\documentclass{article}
\usepackage{amssymb}

\input{tcilatex}

\begin{document}

\title{5D relativistic atom optics and interferometry\\
---------------------------------------------------------}
\author{{\Large Christian J. Bord\'{e}} \\
%EndAName
{SYRTE, Observatoire de Paris, F-75014 Paris, France} \\
and {Laboratoire de Physique des Lasers, F-93430 Villetaneuse, France}\\
{http://christian.j.borde.free.fr}}
\maketitle

\begin{abstract}
This contribution is an update of a previous presentation of 5D matter-wave
optics and interferometry with a correction of some algebraic errors.
Electromagnetic interactions are explicitly added in the 5D metric tensor in
complete analogy with Kaluza's work. The 5D Lagrangian is rederived and an
expression for the Hamiltonian suitable for the parabolic approximation is
presented. The corresponding equations of motion are also given. The 5D
action is shown to cancel for the actual trajectory which is a null
geodesics of the 5D metric. This presentation is mainly devoted to the
classical aspects of the theory and only general consequences for the
quantum phase of matter-waves are outlined. The application to Bord\'{e}%
-Ramsey interferometers is given as an illustration.
\end{abstract}

\section{\protect\bigskip Introduction}

The foundations of relativistic 5D-optics for matter waves have been
presented in an earlier publication \cite{Borde08}. This is a natural
framework to unify and compare photon and atom optics thanks to formulas
valid for arbitrary mass. The concept of mass and its relationship with
proper time in terms of associated dynamical variables and conjugate quantum
observables are presented again here. Gravito-inertial fields and
electromagnetic fields are included in the 5D metric tensor as in Kaluza's
theory. A corrected expression is given for the 5D Lagrangian and
corresponding equations of motion are derived. As in 4D, a superaction makes
the link with the quantum mechanical phase in 5D. The 5D generalization of
the ABCD theorem \cite{Houches,CRAS,metrologia,GRG} for matter-wave packets
leads to a single formula for the quantum phase in presence of external
fields taking into account the internal degrees of freedom of the particle.

\section{The status of mass in classical relativistic mechanics: from 4 to 5
dimensions}

In special relativity, the total energy $E$\ and the momentum components $%
p^{1},p^{2},p^{3}$ \ of a particle, transform as the contravariant
components of a four-vector

\begin{equation}
p^{\mu }=(p^{0},p^{1},p^{2},p^{3})=(E/c,\overrightarrow{p})
\end{equation}%
and the covariant components are given by : 
\begin{equation}
p_{\mu }=g_{\mu \nu }p^{\mu }
\end{equation}%
where $g_{\mu \nu }$ is the metric tensor. In Minkowski space of signature $%
(+,-,-,-)$:%
\begin{equation}
p_{\mu }=(p_{0},p_{1},p_{2},p_{3})=(E/c,-p^{1},-p^{2},-p^{3})
\end{equation}%
These components are conserved quantities when the system considered is
invariant under corresponding space-time translations. They will become the
generators of space-time translations in the quantum theory. For massive
particles of rest mass $m$, they are connected by the following
energy-momentum relation (see figure 1): 
\begin{equation}
E^{2}=p^{2}c^{2}+m^{2}c^{4}
\end{equation}%
or, in manifestly covariant form,%
\begin{equation}
p^{\mu }p_{\mu }-m^{2}c^{2}=0  \label{modulus}
\end{equation}%
This equation cannot be considered as a definition of mass since the origin
of mass is not in the external motion but rather in an internal motion (see
Appendix A). It simply relates two relativistic invariants and gives a
relativistic expression for the total energy. Thus mass appears as an
additional momentum component $mc$ corresponding to internal degrees of
freedom of the object and which adds up quadratically with external
components of the momentum to yield the total energy squared (Pythagoras'
theorem). In the reference frame in which $p=0$ the mass squared is
responsible for the total energy and can thus be seen as stored internal
energy just like kinetic energy is a form of external energy. Even when this
internal energy is purely kinetic e.g. in the case of a photon in a box, it
appears as pure mass $m^{\ast }$ for the global system (i.e. the box). This
new mass is the relativistic mass of the stored particle:%
\begin{equation}
m^{\ast }c^{2}=\sqrt{p^{2}c^{2}+m^{2}c^{4}}
\end{equation}%
The concept of relativistic mass has been criticized in the past but, as we
shall see, it becomes relevant for embedded systems. We may have a hierarchy
of composed objects (e.g. nuclei, atoms, molecules, atomic clock ...) and at
each level the mass $m^{\ast }$ of the larger object is given by the sum of
energies $p^{0}$ of the inner particles. It transforms as $p^{0}$ with the
internal coordinates and is a scalar with respect to the upper level
coordinates.

Mass is conserved when the system under consideration is invariant in a
proper time translation and will become the generator of such translations
in the quantum theory. In the case of atoms, the internal degrees of freedom
give rise to a mass which varies with the internal excitation. For example,
in the presence of an electromagnetic field inducing transitions between
internal energy levels, the mass of atoms becomes time-dependent (Rabi
oscillations). It is thus necessary to enlarge the usual framework of
dynamics to introduce this new dynamical variable as a fifth component of
the energy-momentum vector.

\FRAME{fhFU}{6.1428in}{4.5214in}{0pt}{\Qcb{5D energy-momentum picture}}{}{%
fig1.jpg}{\special{language "Scientific Word";type
"GRAPHIC";maintain-aspect-ratio TRUE;display "USEDEF";valid_file "F";width
6.1428in;height 4.5214in;depth 0pt;original-width 8.5263in;original-height
6.2656in;cropleft "0";croptop "1";cropright "1";cropbottom "0";filename
'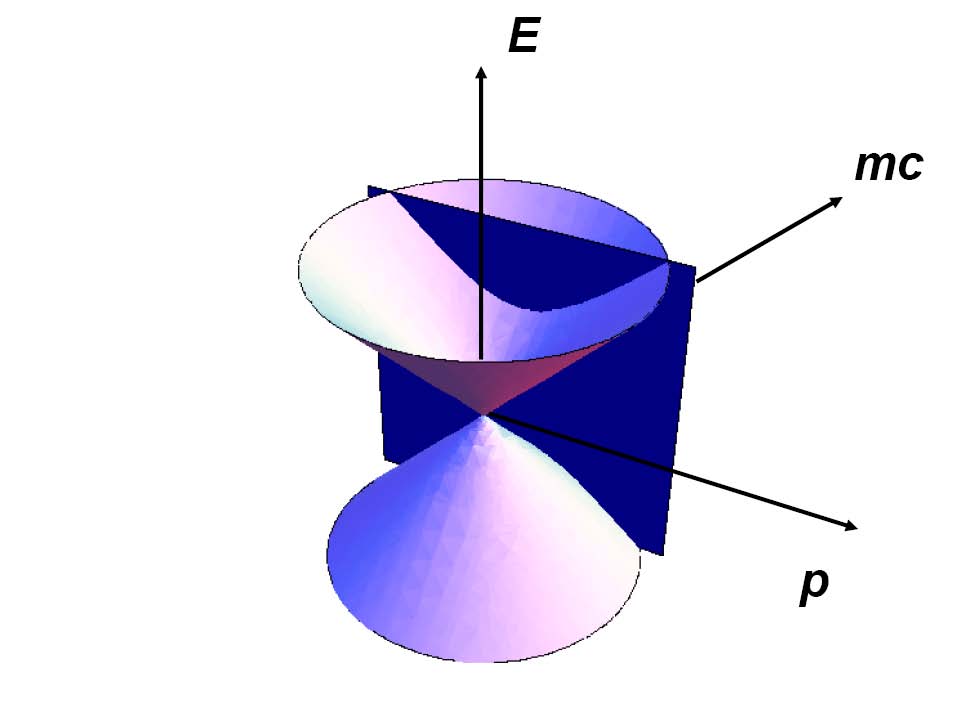';file-properties "XNPEU";}}

\bigskip Equation (\ref{modulus}) can be written with a five dimensional
notation :%
\begin{equation}
G^{\hat{\mu}\hat{\nu}}\widehat{p}_{\hat{\mu}}\widehat{p}_{\hat{\nu}}=0\text{
with }\hat{\mu},\hat{\nu}=0,1,2,3,4
\end{equation}%
where $\widehat{p}_{\hat{\mu}}=(p_{\mu },p_{4}=-mc)$ ;\ \ $G^{\mu \nu
}=g^{\mu \nu }$\ ;\ \ $G^{\hat{\mu}4}=G^{4\hat{\nu}}=0$ \ ;\ \ $G^{44}=\
G_{44}=-1$

This leads us to consider also the picture in the coordinate space and its
extension to five dimensions. As in the previous case, we have a four-vector
representing the space-time position of a particle:%
\[
x^{\mu }=(ct,x,y,z) 
\]%
and in view of the extension to general relativity:%
\begin{equation}
dx^{\mu }=(cdt,dx,dy,dz)=(dx^{0},dx^{1},dx^{2},dx^{3})
\end{equation}

The relativistic invariant is, in this case, the elementary interval $ds$,
also expressed with the proper time $\tau $\ of the particle:%
\begin{equation}
ds^{2}=dx^{\mu }dx_{\mu }=c^{2}dt^{2}-d\overrightarrow{x}^{2}=c^{2}d\tau ^{2}
\label{interval}
\end{equation}%
which is, as that was already the case for mass, equal to zero for light

\begin{equation}
ds^{2}=0
\end{equation}%
and this defines the usual light cone in space-time.

For massive particles proper time and interval are non-zero and equation (%
\ref{interval}) defines again an hyperboloid. As in the energy-momentum
picture we may enlarge our space-time with the additional dimension \ $%
s=c\tau $

\begin{equation}
d\widehat{x}^{\widehat{\mu }}=(cdt,dx,dy,dz,cd\tau
)=(dx^{0},dx^{1},dx^{2},dx^{3},dx^{4})
\end{equation}%
and introduce a generalized light cone for massive particles\footnote{%
In this picture, anti-particles have a negative mass and propagate backwards
on the fifth axis as first pointed out by Feynman. Still, their relavistic
mass $m^{\ast }$ is positive and hence they follow the same trajectories as
particles in gravitational fields as we shall see from the equations of
motion.}

\begin{equation}
d\sigma ^{2}=G_{^{\hat{\mu}\hat{\nu}}}d\widehat{x}^{\hat{\mu}}d\widehat{x}^{%
\widehat{\nu }}=c^{2}dt^{2}-d\overrightarrow{x}^{2}-c^{2}d\tau ^{2}=0
\end{equation}%
As pointed out in the case of mass, proper time is not defined by this
equation from other coordinates but is rather a true evolution parameter
representative of the internal evolution of the object. It coincides
numerically with the time coordinate in the frame of the object through the
relation:

\begin{equation}
cd\tau =\sqrt{G_{00}}dx^{0}
\end{equation}%
\FRAME{ftbpFU}{6.1428in}{4.5214in}{0pt}{\Qcb{ 5D coordinates}}{}{fig2.jpg}{%
\special{language "Scientific Word";type "GRAPHIC";maintain-aspect-ratio
TRUE;display "USEDEF";valid_file "F";width 6.1428in;height 4.5214in;depth
0pt;original-width 8.5263in;original-height 6.2656in;cropleft "0";croptop
"1";cropright "1";cropbottom "0";filename '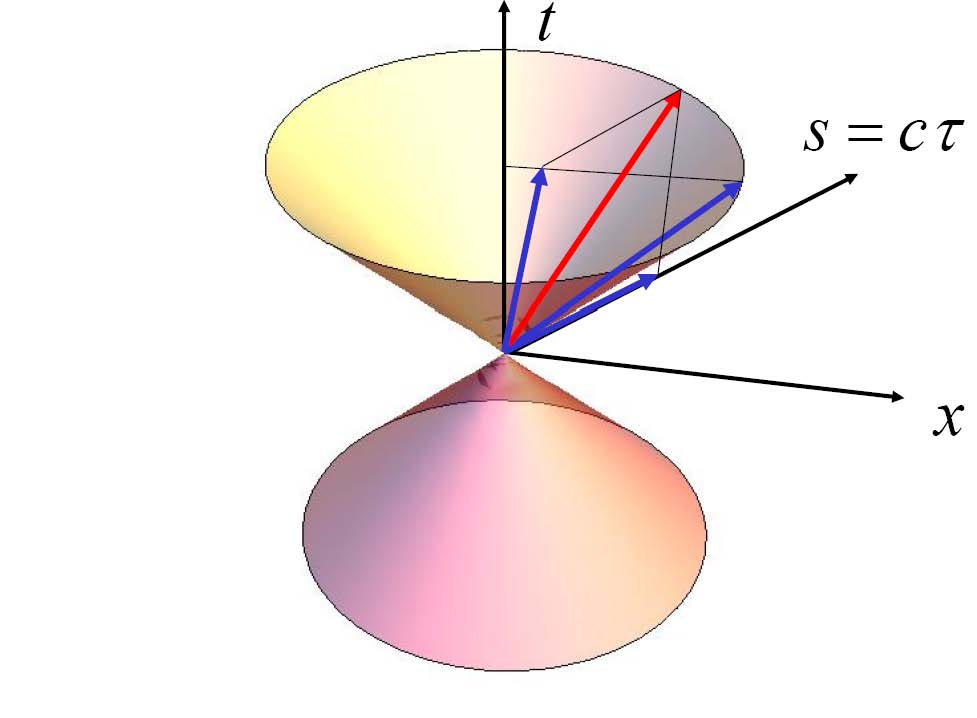';file-properties
"XNPEU";}}

\bigskip Finally, if we combine momenta and coordinates to form a mixed
scalar product, we obtain a new relativistic invariant which is the
differential of the action. In 4D:

\begin{equation}
dS=-p_{\mu }dx^{\mu }
\end{equation}%
and in 5D we shall therefore introduce the superaction:%
\begin{equation}
\widehat{S}=-\int \widehat{p}_{\hat{\mu}}d\widehat{x}^{\hat{\mu}}
\label{Superaction}
\end{equation}%
equivalent to%
\begin{equation}
\widehat{p}_{\hat{\mu}}=-\frac{\partial \widehat{S}}{\partial \widehat{x}^{%
\hat{\mu}}}\text{ \ with }\hat{\mu}=0,1,2,3,4
\end{equation}%
If this is substituted in%
\begin{equation}
G^{\hat{\mu}\hat{\nu}}\widehat{p}_{\hat{\mu}}\widehat{p}_{\hat{\mu}}=0\text{ 
}
\end{equation}%
we obtain the Hamilton-Jacobi equation in 5D%
\begin{equation}
G^{\hat{\mu}\hat{\nu}}\partial _{\hat{\mu}}\widehat{S}\partial _{\hat{\nu}}%
\widehat{S}=0\text{ }
\end{equation}%
which has the same form as the eikonal equation for light in 4D. It is
already this striking analogy which pushed Louis de Broglie to identify
action and the phase of a matter wave in the 4D case. We shall follow the
same track for a quantum approach in our 5D case.

What is the link between the three previous invariants given above? As in
optics, the direction of propagation of a particle is determined by the
momentum vector tangent to the trajectory. The 5D momentum can therefore be
written in the form:%
\begin{equation}
\widehat{p}^{\hat{\mu}}=d\widehat{x}^{\hat{\mu}}/d\lambda
\end{equation}%
where $\lambda $ is an affine parameter varying along the ray. This is
consistent with the invariance of these quantities for uniform motion.

In 4D the canonical 4-momentum is: $\ \ $%
\begin{equation}
p_{\mu }=mc\frac{g_{\mu \nu }dx^{\nu }}{\sqrt{g_{\mu \nu }dx^{\mu }dx^{\nu }}%
}=mcg_{\mu \nu }u^{\nu }
\end{equation}%
where $u^{\nu }=dx^{\nu }/d\tau $ is the normalized 4-velocity with $d\tau =%
\sqrt{g_{\mu \nu }dx^{\mu }dx^{\nu }}$ given by (\ref{interval}).

We observe that $d\lambda $ can always be written as the ratio of a time to
a mass:%
\begin{equation}
d\lambda =\frac{d\tau }{m}=\frac{dt}{m^{\ast }}=\frac{d\theta }{M}=...
\end{equation}%
where\ $\tau $ is the proper time of individual particles (e.g. atoms in a
clock or in a molecule), $t$ is the time coordinate of the composed object
(clock, interferometer or molecule) and $\theta $ its proper time; $%
m,m^{\ast },M$ are respectively the mass, the relativistic mass of
individual particles and their contribution to the scalar mass of the device
or composed object.

In the usual paradigm of relativity, the time $t$ is a coordinate variable
and the proper time $\tau $ is taken as the evolution parameter to describe
the motion of particles in space-time. In this presentation however, proper
time is an independent coordinate describing the internal motion of massive
particles, so that we shall rather chose the coordinate time as the
evolution parameter. Another good reason for this choice is that, in order
to describe an ensemble of atoms or of atom waves within a clock or an atom
interferometer, it cannot be a good choice to use the proper time of a
specific atom to describe the full device. We shall therefore write in 5D:%
\begin{equation}
\widehat{p}_{\hat{\mu}}=m^{\ast }G_{\hat{\mu}\hat{\nu}}\dot{\widehat{x}}^{%
\widehat{\nu }}=m^{\ast }\dot{\widehat{x}}_{\hat{\mu}}
\end{equation}%
expressed with the "relativistic mass" :%
\begin{equation}
m^{\ast }=m\frac{dt}{d\tau }=\frac{mc}{\sqrt{g_{\mu \nu }\dot{x}^{\mu }\dot{x%
}^{\nu }}}
\end{equation}%
and where the dot refers to derivation with respect to a "laboratory time"
(identical to the proper time $\theta $\ of the apparatus only in the
absence of gravitation or inertial effects). With this choice $\dot{\widehat{%
x}}^{0}=c$ and $\widehat{p}^{0}=m^{\ast }c.$ An alternate choice could be to
take the proper time $\theta $ of the full device as the evolution
parameter. In which case:%
\begin{equation}
cd\theta =\sqrt{G_{00}}dx^{0}\text{ and }M=m^{\ast }\sqrt{G_{00}}
\end{equation}%
From :%
\begin{equation}
d\sigma ^{2}=G_{^{\hat{\mu}\hat{\nu}}}d\widehat{x}^{\hat{\mu}}d\widehat{x}^{%
\widehat{\nu }}=0
\end{equation}%
we infer in 5D%
\begin{equation}
d\widehat{S}=0
\end{equation}%
and in 4D%
\begin{equation}
dS=-p_{\mu }dx^{\mu }=-mc^{2}d\tau
\end{equation}%
In Appendix A, we generalize these relations to an object, such as a clock,
a molecule.., composed of a number of subparticles and illustrate the origin
of proper time as coming from the inner structure of the object.

\section{\protect\bigskip Generalization in the presence of gravitational
and electromagnetic interactions}

The previous 5D scheme can be extended to general relativity with a 4D
metric tensor $g^{\mu \nu }$ and an electromagnetic 4-potential $\ A_{\mu }$

\bigskip 
\begin{equation}
g^{\mu \nu }\left( p_{\mu }-qA_{\mu }\right) \left( p_{\nu }-qA_{\nu
}\right) =m^{2}c^{2}  \label{KGA}
\end{equation}

($q=-e$ for the electron).

We shall search for a metric tensor $G_{\hat{\mu}\hat{\nu}}$ for 5D such
that the generalized interval given by: 
\[
d\sigma ^{2}=G_{\hat{\mu}\hat{\nu}}d\widehat{x}^{\hat{\mu}}d\widehat{x}^{%
\widehat{\nu }} 
\]%
is an invariant.

Let us recall that, from the equivalence principle, the metric tensor $%
g^{\mu \nu }$ can be obtained from the Minkovski flat space-time tensor $%
\eta ^{\mu \nu }$ using infinitesimal frame transformations from a locally
inertial frame. Quite generally any infinitesimal coordinate transformation
considered as a gauge transformation can be used to introduce a component of
the gravito-inertial field. As an example, in 4D, the transformation (case
of a rotation):

\bigskip 
\begin{eqnarray}
dx^{\prime i} &=&dx^{i}+\alpha _{0}^{i}dx^{0}  \nonumber \\
dx^{\prime 0} &=&dx^{0}
\end{eqnarray}%
transforms the interval%
\begin{equation}
ds^{2}=g_{00}^{\prime }(dx^{\prime 0})^{2}+g_{ij}^{\prime }dx^{\prime
i}dx^{\prime j}
\end{equation}%
into%
\begin{equation}
ds^{2}=g_{00}(dx^{0})^{2}+2g_{0i}dx^{0}dx^{i}+g_{ij}dx^{i}dx^{j}
\end{equation}%
with%
\begin{eqnarray}
g_{00} &=&g_{00}^{\prime }+\alpha _{0}^{i}\alpha _{0}^{j}g_{ij}^{\prime } \\
g_{0i} &=&\alpha _{0}^{i}g_{ij}^{\prime } \\
g_{ij} &=&g_{ij}^{\prime }
\end{eqnarray}

\begin{eqnarray}
g^{00} &=&g^{^{\prime }00}=1/g_{00}^{\prime } \\
g^{^{\prime }ij} &=&1/g_{ij}^{\prime }
\end{eqnarray}%
Using :

\begin{equation}
g_{ij}g^{i0}=-g^{00}g_{j0}
\end{equation}%
we find%
\begin{eqnarray}
\alpha _{0}^{i} &=&-\frac{g^{i0}}{g^{00}} \\
\alpha _{0}^{i}\alpha _{0}^{j}g_{ij}^{\prime } &=&-\frac{g_{i0}g^{i0}}{g^{00}%
}
\end{eqnarray}%
In the case of rotation we recover the usual metric tensor in the rotating
frame.

The action $S$ becomes%
\begin{eqnarray}
S &=&-\int p_{\mu }^{\prime }dx^{\prime \mu }=-\int p_{0}^{\prime
}dx^{\prime 0}-\int p_{i}^{\prime }dx^{\prime i} \\
&=&-\int p_{0}^{\prime }dx^{0}-\int p_{i}^{\prime }(dx^{i}+\alpha
_{0}^{i}dx^{0})
\end{eqnarray}

\[
S=-\int \left( p_{0}^{\prime }+p_{i}^{\prime }\alpha _{0}^{i}\right)
dx^{0}-\int p_{i}^{\prime }dx^{i}=-\int p_{\mu }dx^{\mu } 
\]%
which gives the Sagnac phase as $\int \left( p_{i}g^{i0}/g^{00}\right)
dx^{0} $.

The same approach can be used with the fifth dimension by introducing the
gauge transformation%
\begin{eqnarray}
dx^{\prime 4} &=&dx^{4}+\beta _{\mu }^{4}d\widehat{x}^{\mu }  \nonumber \\
d\widehat{x}^{\prime \mu } &=&d\widehat{x}^{\mu }
\end{eqnarray}%
to generate the off-diagonal elements $G_{\mu 4}$%
\[
d\sigma ^{2}=G_{44}\left( dx^{4}\right) ^{2}+2G_{44}\beta _{\mu }^{4}dx^{4}d%
\widehat{x}^{\mu }+\left( g_{\mu \nu }+\beta _{\mu }^{4}\beta _{\nu
}^{4}G_{44}\right) d\widehat{x}^{\mu }d\widehat{x}^{\nu } 
\]

\begin{eqnarray}
G_{44} &=&G_{44}^{\prime } \\
G_{\mu 4} &=&\beta _{\mu }^{4}G_{44} \\
G_{\mu \nu } &=&g_{\mu \nu }+\beta _{\mu }^{4}\beta _{\nu }^{4}G_{44}
\end{eqnarray}%
The superaction $\widehat{S}$ given by (\ref{Superaction}) becomes%
\begin{eqnarray}
\widehat{S} &=&-\int \widehat{p}_{\hat{\mu}}d\widehat{x}^{\prime \hat{\mu}%
}=-\int p_{\mu }d\widehat{x}^{\prime \mu }-\int \widehat{p}_{4}dx^{\prime 4}
\\
&=&-\int p_{\mu }d\widehat{x}^{\mu }+\int mc(dx^{4}+\beta _{\mu }^{4}d%
\widehat{x}^{\mu })
\end{eqnarray}

\begin{equation}
\widehat{S}=-\int \left( p_{\mu }-mc\beta _{\mu }^{4}\right) d\widehat{x}%
^{\mu }+\int mc^{2}d\tau
\end{equation}%
which yields the Aharonov-Bohm phase if $mc\beta _{\mu }^{4}=qA_{\mu }$.

The metric tensor in five dimensions $G_{\mu \nu }$ is thus written as in
Kaluza's theory to include the electromagnetic gauge field potential $A_{\mu
}$ 
\begin{eqnarray}
G_{\hat{\mu}\hat{\nu}} &=&\left( 
\begin{array}{cc}
G_{\mu \nu } & G_{\mu 4} \\ 
G_{4\nu } & G_{44}%
\end{array}%
\right) =\left( 
\begin{array}{cc}
g_{\mu \nu }+\kappa ^{2}G_{44}A_{\mu }A_{\nu } & \kappa G_{44}A_{\mu } \\ 
\kappa G_{44}A_{\nu } & G_{44}%
\end{array}%
\right)  \nonumber \\
G^{\hat{\mu}\hat{\nu}} &=&\left( 
\begin{array}{cc}
G^{\mu \nu } & G^{\mu 4} \\ 
G^{4\nu } & G^{44}%
\end{array}%
\right) =\left( 
\begin{array}{cc}
g^{\mu \nu } & -\kappa A^{\mu } \\ 
-\kappa A^{\nu } & G^{44}%
\end{array}%
\right)
\end{eqnarray}%
where $\kappa $\ is given by the gyromagnetic ratio of the object. This
metric tensor is such that

\bigskip 
\begin{eqnarray}
G^{\hat{\mu}\widehat{\lambda }}G_{\widehat{\lambda }\hat{\nu}} &=&\left( 
\begin{array}{cc}
G^{\mu \lambda } & G^{\mu 4} \\ 
G^{4\lambda } & G^{44}%
\end{array}%
\right) \left( 
\begin{array}{cc}
G_{\lambda \nu } & G_{\lambda 4} \\ 
G_{4\nu } & G_{44}%
\end{array}%
\right) =\delta _{\hat{\nu}}^{\hat{\mu}} \\
&=&\left( 
\begin{array}{cc}
G^{\mu \lambda } & -\kappa A^{\mu } \\ 
-\kappa A^{\lambda } & G^{44}%
\end{array}%
\right) \left( 
\begin{array}{cc}
g_{\lambda \nu }+\kappa ^{2}G_{44}A_{\lambda }A_{\nu } & +\kappa
G_{44}A_{\lambda } \\ 
+\kappa G_{44}A_{\nu } & G_{44}%
\end{array}%
\right)   \nonumber \\
&=&\left( 
\begin{array}{cc}
G^{\mu \lambda }g_{\lambda \nu } & \kappa G_{44}G^{\mu \lambda }A_{\lambda
}-\kappa G_{44}A^{\mu }=0 \\ 
-\kappa A^{\lambda }(g_{\lambda \nu }+\kappa ^{2}G_{44}A_{\lambda }A_{\nu
})+\kappa G_{44}G^{44}A_{\nu } & -\kappa ^{2}G_{44}A^{\lambda }A_{\lambda
}+G^{44}G_{44}%
\end{array}%
\right) =\delta _{\hat{\nu}}^{\hat{\mu}}  \nonumber
\end{eqnarray}%
which implies%
\begin{eqnarray}
G^{\mu \lambda }g_{\lambda \nu } &=&\delta _{\nu }^{\mu }\text{ \ \ } 
\nonumber \\
G^{44} &=&1/G_{44}+\kappa ^{2}A^{\lambda }A_{\lambda }
\end{eqnarray}%
\bigskip The equation :%
\begin{equation}
G^{\hat{\mu}\hat{\nu}}\widehat{p}_{\hat{\mu}}\widehat{p}_{\hat{\nu}}=0\text{ 
}
\end{equation}%
with%
\begin{equation}
\widehat{p}_{\hat{\mu}}=(p_{\mu },-mc)
\end{equation}%
and \ $G_{44}=-1$ is therefore equivalent to equation (\ref{KGA})%
\begin{equation}
g^{\mu \nu }\left( p_{\mu }-qA_{\mu }\right) \left( p_{\nu }-qA_{\nu
}\right) =m^{2}c^{2}
\end{equation}%
Higher order electromagnetic interactions are introduced via the multipolar
expansion$\ \ \ \ p_{\mu }-qA_{\mu }+Q^{\lambda }F_{\mu \lambda }$, where
dipole moments will become operators in the quantum description.

\section{\protect\bigskip Hamiltonian and Lagrangian : parabolic
approximation}

In some cases it is convenient to assume that the energy $E$ is close to a
known value $E_{0}$ either because energy is conserved and remains equal to
its initial value or because of a slow variation of parameters. This means
that the usual hyperbolic dispersion curve is locally approximated by the
parabola tangent to the hyperbola for the energy $E_{0}$. This approximation
scheme applies to massive as well as to massless particles. We can then make
use of the identity: $E=\frac{E_{0}}{2}+\frac{E^{2}}{2E_{0}}+O(\varepsilon
^{2})$ valid to second-order in $\varepsilon =E-E_{0}$ (parabolic
approximation).

Let us start with the exact formula:

\begin{equation}
\widehat{p}^{0}=\frac{\widehat{p}^{0}}{2}+\frac{(\widehat{p}^{0})^{2}}{2%
\widehat{p}^{0}}
\end{equation}%
in which $(\widehat{p}^{0})^{2}$ is obtained from: 
\begin{eqnarray}
0 &=&G^{\hat{\mu}\hat{\nu}}\widehat{p}_{\hat{\mu}}\widehat{p}_{\hat{\nu}}%
\text{ }=G^{00}(\widehat{p}_{0})^{2}+2G^{0\widehat{i}}\widehat{p}_{0}%
\widehat{p}_{\widehat{i}}+G^{\widehat{i}\widehat{j}}\widehat{p}_{\widehat{i}}%
\widehat{p}_{\widehat{j}} \\
&=&G^{00}(\widehat{p}_{0}+\frac{G^{0\widehat{i}}}{G^{00}}\widehat{p}_{%
\widehat{i}})^{2}+\left( G^{\widehat{i}\widehat{j}}-\frac{G^{0\widehat{i}%
}G^{0\widehat{j}}}{G^{00}}\right) \widehat{p}_{\widehat{i}}\widehat{p}_{%
\widehat{j}} \\
&=&\frac{1}{G^{00}}(\widehat{p}^{0})^{2}+\hat{f}^{\widehat{i}\widehat{j}}%
\widehat{p}_{\widehat{i}}\widehat{p}_{\widehat{j}}
\end{eqnarray}%
i.e.:%
\begin{equation}
(\widehat{p}^{0})^{2}=-G^{00}\hat{f}^{\widehat{i}\widehat{j}}\widehat{p}_{%
\widehat{i}}\widehat{p}_{\widehat{j}}
\end{equation}%
where 
\begin{equation}
\hat{f}^{\widehat{i}\widehat{j}}=G^{\widehat{i}\widehat{j}}-\frac{G^{0%
\widehat{i}}G^{0\widehat{j}}}{G^{00}}  \label{H5}
\end{equation}%
is the 4D metric tensor, inverse of $G_{\widehat{i}\widehat{j}}$. Hence%
\begin{equation}
\widehat{p}^{0}=\frac{\widehat{p}^{0}}{2}-\frac{G^{00}\hat{f}^{\widehat{i}%
\widehat{j}}\widehat{p}_{\widehat{i}}\widehat{p}_{\widehat{j}}}{2\widehat{p}%
^{0}}=G^{00}\widehat{p}_{0}+G^{0\widehat{i}}\widehat{p}_{\widehat{i}}
\end{equation}%
and%
\begin{eqnarray}
\widehat{p}_{0} &=&\frac{\widehat{p}^{0}}{2G^{00}}-\frac{\hat{f}^{\widehat{i}%
\widehat{j}}\widehat{p}_{\widehat{i}}\widehat{p}_{\widehat{j}}}{2\widehat{p}%
^{0}}-\frac{G^{0\widehat{j}}\widehat{p}_{\widehat{j}}c}{G^{00}} \\
\widehat{i},\widehat{j} &=&1,2,3,4  \nonumber
\end{eqnarray}%
With the choice of time coordinate such that$\dot{\text{ }\widehat{x}}^{0}=c$
the Hamiltonian can be finally written: 
\begin{eqnarray}
H &=&\frac{m^{\ast }c^{2}}{2G^{00}}-\frac{\hat{f}^{\widehat{i}\widehat{j}}%
\widehat{p}_{\widehat{i}}\widehat{p}_{\widehat{j}}}{2m^{\ast }}-\frac{G^{0%
\widehat{j}}\widehat{p}_{\widehat{j}}c}{G^{00}}  \label{H4} \\
\widehat{i},\widehat{j} &=&1,2,3,4  \nonumber
\end{eqnarray}%
This expression is exact but requires the knowledge of the relativistic mass 
$m^{\ast }$. In the parabolic approximation this quantity will finally be
approximated by its central value. From the previous exact expression of the
Hamiltonian, the Lagrangian is recovered as:%
\begin{equation}
\widehat{L}=-\widehat{p}_{\hat{\mu}}\dot{\widehat{x}}^{\hat{\mu}}=-\frac{1}{2%
}m^{\ast }G_{\hat{\mu}\hat{\nu}}\dot{\widehat{x}}^{\hat{\mu}}\dot{\widehat{x}%
}^{\hat{\nu}}
\end{equation}

\section{Equations of motion}

From this Lagrangian we may infer the following equations of motion:

\begin{equation}
\widehat{p}_{\hat{\mu}}=-\frac{\partial \widehat{L}}{\partial \dot{\widehat{x%
}}^{\hat{\mu}}}=m^{\ast }G_{\hat{\mu}\hat{\nu}}\dot{\widehat{x}}^{\hat{\nu}}
\end{equation}%
i.e.%
\begin{equation}
\dot{\widehat{x}}^{\widehat{i}}=\frac{\hat{f}^{\widehat{i}\widehat{j}}%
\widehat{p}_{\widehat{j}}}{m^{\ast }}+\frac{G^{0\widehat{j}}c}{G^{00}}
\end{equation}%
and 
\begin{equation}
\dot{\widehat{p}}_{\widehat{\lambda }}=\frac{1}{2}m^{\ast }\left( \partial _{%
\widehat{\lambda }}G_{\hat{\mu}\hat{\nu}}\right) \dot{\widehat{x}}^{\hat{\mu}%
}\dot{\widehat{x}}^{\widehat{\nu }}
\end{equation}%
or%
\begin{equation}
\dot{\widehat{p}}^{\hat{\mu}}=\frac{1}{2}m^{\ast }G^{\hat{\mu}\widehat{%
\lambda }}\left( \partial _{\widehat{\lambda }}G_{\widehat{\kappa }\hat{\nu}%
}-2\partial _{\widehat{\nu }}G_{\widehat{\lambda }\widehat{\kappa }}\right) 
\dot{\widehat{x}}^{\widehat{\kappa }}\dot{\widehat{x}}^{\widehat{\nu }}
\end{equation}%
These equations can be compared to those obtained either from the equation
for geodesic lines in 5D obtained from $\delta d\sigma ^{2}=0$ with%
\begin{equation}
d\sigma ^{2}=G_{\hat{\mu}\hat{\nu}}d\widehat{x}^{\hat{\mu}}d\widehat{x}^{%
\hat{\nu}}
\end{equation}%
or from the condition:%
\begin{equation}
D\dot{\widehat{x}}^{\hat{\mu}}=0
\end{equation}%
We proceed as in 4D and find:%
\[
\ddot{\widehat{x}}^{\hat{\mu}}+\text{ }^{(5)}\Gamma _{\widehat{\nu }\widehat{%
\lambda }}^{\hat{\mu}}\dot{\widehat{x}}^{\widehat{\nu }}\dot{\widehat{x}}^{%
\widehat{\lambda }}=0 
\]%
with%
\begin{equation}
^{(5)}\Gamma _{\widehat{\nu }\widehat{\lambda }}^{\hat{\mu}}\dot{\widehat{x}}%
^{\widehat{\nu }}\dot{\widehat{x}}^{\widehat{\lambda }}=\frac{1}{2}G^{\hat{%
\mu}\widehat{\kappa }}\left( 2\partial _{\widehat{\lambda }}G_{\widehat{%
\kappa }\widehat{\nu }}-\partial _{\widehat{\kappa }}G_{\widehat{\nu }%
\widehat{\lambda }}\right) \dot{\widehat{x}}^{\widehat{\nu }}\dot{\widehat{x}%
}^{\widehat{\lambda }}
\end{equation}%
We wish now to check that we recover the usual equations of motion in 4D
when the metric is independent of the 5th \ coordinate:%
\begin{equation}
\ddot{\widehat{x}}^{\mu }+^{(5)}\Gamma _{\nu \lambda }^{\mu }\dot{\widehat{x}%
}^{\nu }\dot{\widehat{x}}^{\lambda }+^{(5)}\Gamma _{4\lambda }^{\mu }\dot{%
\widehat{x}}^{4}\dot{\widehat{x}}^{\lambda }+^{(5)}\Gamma _{\nu 4}^{\mu }%
\dot{\widehat{x}}^{4}\dot{\widehat{x}}^{\nu }+^{(5)}\Gamma _{44}^{\mu }\dot{%
\widehat{x}}^{4}\dot{\widehat{x}}^{4}=0
\end{equation}%
with%
\begin{eqnarray}
^{(5)}\Gamma _{4\lambda }^{\mu } &=&\frac{G_{44}}{2}\kappa F_{\lambda }^{\mu
} \\
^{(5)}\Gamma _{\nu 4}^{\mu } &=&\frac{G_{44}}{2}\kappa F_{\nu }^{\mu } \\
^{(5)}\Gamma _{44}^{\mu } &=&G^{\mu \nu }\partial _{4}G_{\nu 4}=0
\end{eqnarray}%
The Christoffel symbols in 4D and 5D are connected by:

\begin{equation}
\text{ }^{(5)}\Gamma _{\nu \lambda }^{\mu }-\text{ }^{(4)}\Gamma _{\nu
\lambda }^{\mu }=\frac{\kappa ^{2}}{2}\left( A_{\nu }F_{\lambda }^{\mu
}+A_{\lambda }F_{\nu }^{\mu }\right)
\end{equation}%
Hence%
\begin{equation}
\ddot{\widehat{x}}^{\mu }+\text{ }^{(4)}\Gamma _{\nu \lambda }^{\mu }\dot{%
\widehat{x}}^{\nu }\dot{\widehat{x}}^{\lambda }+\frac{\kappa ^{2}}{2}\left(
A_{\nu }F_{\lambda }^{\mu }+A_{\lambda }F_{\nu }^{\mu }\right) \dot{\widehat{%
x}}^{\nu }\dot{\widehat{x}}^{\lambda }=-G_{44}\dot{\widehat{x}}^{4}\kappa
F_{\lambda }^{\mu }\dot{\widehat{x}}^{\lambda }
\end{equation}%
using%
\begin{equation}
-G_{44}\dot{\widehat{x}}^{4}=-\dot{\widehat{x}}_{4}+G_{4\lambda }\dot{%
\widehat{x}}^{\widehat{\lambda }}
\end{equation}%
we recover the usual 4D equation of motion:%
\begin{equation}
m^{\ast }(\ddot{\widehat{x}}^{\mu }+\text{ }^{(4)}\Gamma _{\nu \lambda
}^{\mu }\dot{\widehat{x}}^{\nu }\dot{\widehat{x}}^{\lambda })=qF_{\lambda
}^{\mu }\dot{\widehat{x}}^{\lambda }
\end{equation}%
since%
\begin{equation}
m^{\ast }\dot{\widehat{x}}_{4}=\widehat{p}_{4}=-mc
\end{equation}%
If we reintroduce the dependence in the fifth coordinate, we obtain the rate
of mass change associated with the change of internal motion induced by an
electromagnetic field:%
\begin{eqnarray}
\dot{\widehat{p}}_{4} &=&\frac{1}{2}m^{\ast }\left( \partial _{4}G_{\hat{\mu}%
\hat{\nu}}\right) \dot{\widehat{x}}^{\hat{\mu}}\dot{\widehat{x}}^{\widehat{%
\nu }}  \nonumber \\
&=&m^{\ast }\left( \partial _{4}G_{\mu 4}\right) \dot{\widehat{x}}^{\mu }%
\dot{\widehat{x}}^{4}
\end{eqnarray}%
and similar expressions for the rate of energy-momentum changes induced by
internal transitions. In the case of electric dipole transitions, the photon
energy-momentum is exchanged at the Rabi frequency rate. However in this
approximation we do not obtain the Rabi oscillations (pendell\"{o}sung)
which require to introduce two coupled modes and therefore a quantum
treatment of their amplitudes.

\section{5D expression of the phase shift}

The total phase difference between both arms of an interferometer is usually
calculated as the sum of three terms: the difference in the action integral
along each path, the difference in the phases imprinted on the atom waves by
the beam splitters and a contribution coming from the splitting of the wave
packets at the exit of the interferometer \cite{CRAS,GRG}. If $\alpha $ and $%
\beta $ are the two branches of the interferometer:\FRAME{ftbpF}{4.1204in}{%
3.0926in}{0pt}{}{}{phase2chemins-b.jpg}{\special{language "Scientific
Word";type "GRAPHIC";maintain-aspect-ratio TRUE;display "USEDEF";valid_file
"F";width 4.1204in;height 3.0926in;depth 0pt;original-width
9.5998in;original-height 7.1996in;cropleft "0";croptop "1";cropright
"1";cropbottom "0";filename
'../BadHonnefBordé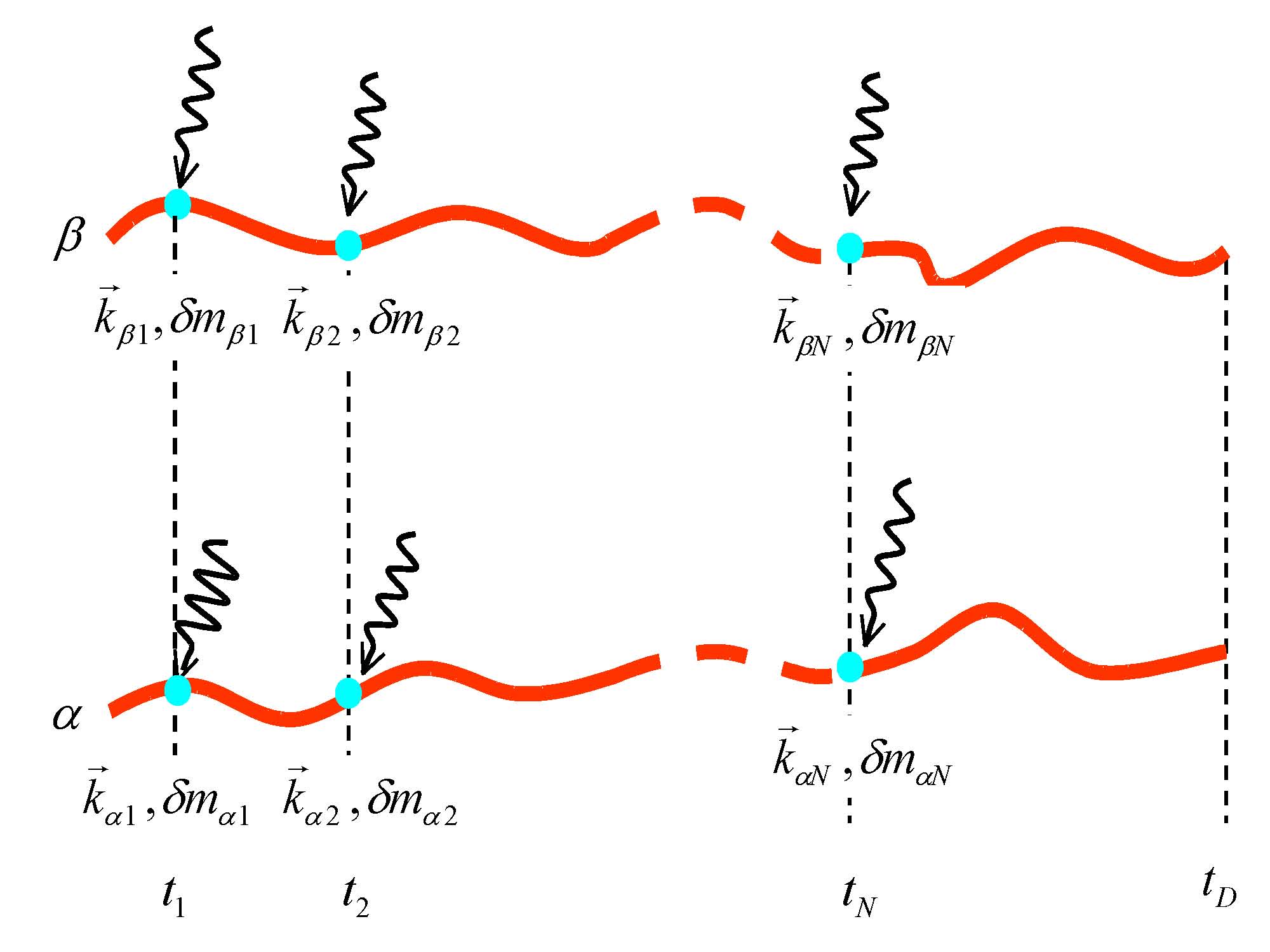';file-properties "XNPEU";}}%
\begin{eqnarray}
\delta \phi (q) &=&\sum_{j=1}^{N}\left[ S_{\beta }\left(
t_{j+1},t_{j}\right) -S_{\alpha }\left( t_{j+1},t_{j}\right) \right] /\hbar 
\nonumber \\
&&+\sum_{j=1}^{N}\left( \tilde{k}_{\beta j}q_{\beta j}-\tilde{k}_{\alpha
j}q_{\alpha j}\right) -\left( \omega _{\beta j}-\omega _{\alpha j}\right)
t_{j}+\left( \varphi _{\beta j}-\varphi _{\alpha j}\right)  \nonumber \\
&&+\left[ \tilde{p}_{\beta ,D}\left( q-q_{\beta ,D}\right) -\tilde{p}%
_{\alpha ,D}\left( q-q_{\alpha ,D}\right) \right] /\hbar
\end{eqnarray}%
where $S_{\alpha j}=S_{\alpha }\left( t_{j+1},t_{j}\right) $ and $S_{\beta
j}=S_{\beta }\left( t_{j+1},t_{j}\right) $ are the action integrals along $%
\alpha $ $(\beta )$ paths; $\hbar k_{\alpha j}(\hbar k_{\beta j})$ are the
momenta transferred to the atoms by the j-th beam splitter along the $\alpha 
$ $(\beta )$ arm; $q_{\alpha j}$ and $q_{\beta j}$ are the classical
coordinates of the centers of the beam splitter/atom interactions; $\omega
_{\alpha j}(\omega _{\beta j})$ are the angular frequencies of the e.m.
waves; $\varphi _{\alpha j}(\varphi _{\beta j})$ are the fixed phases of the
j-th beam splitters; $D$ is the detection port.

With our new approach in 5D the action terms should be replaced by the phase
jumps induced by the beam splitters along the fourth space coordinate $c\tau 
$:%
\begin{equation}
\sum_{j=1}^{N}c^{2}\left[ \delta m_{\beta j}\tau _{\beta j}-\delta m_{\alpha
j}\tau _{\alpha j}\right] /\hbar
\end{equation}%
in which $\delta m_{\beta j}$\ $(\delta m_{\alpha j})$ are the mass changes
introduced by each splitter. To obtain this result we write the action terms
as: 
\begin{equation}
\sum_{j=1}^{N\ }S_{\beta }\left( t_{j+1},t_{j}\right) =\sum_{j=1}^{N\
}-c^{2} \left[ m_{\beta j+1}\tau _{\beta j+1}-(m_{\beta j}+\delta m_{\beta
j})\tau _{\beta j}\right]
\end{equation}%
with $m_{\beta N+1}=m_{\beta D\text{ \ \ \ }}$ and $\tau _{\beta N+1}=\tau
_{\beta D}$. We shift $j$ by one unit for the first term: 
\begin{equation}
\sum_{j=1}^{N\ }S_{\beta }\left( t_{j+1},t_{j}\right) =c^{2}m_{\beta 1}\tau
_{\beta 1}+\sum_{j=1}^{N\ }-c^{2}\left[ m_{\beta j}\tau _{\beta j}-(m_{\beta
j}+\delta m_{\beta j})\tau _{\beta j}\right] -c^{2}m_{\beta D}\tau _{\beta D}
\end{equation}%
We suppress the first term $c^{2}m_{\beta 1}\tau _{\beta 1}$ and add a
current term $c^{2}m_{\beta D}\tau $ following the logic of a phase term
analogous to the spatial terms (these terms are generally eliminated between
both arms of the interferometer but they are indeed new phases arising in
the 5D approach):

\begin{eqnarray}
&&\sum_{j=1}^{N\ }S_{\beta }\left( t_{j+1},t_{j}\right) \text{ is replaced
by }\sum_{j=1}^{N\ }-c^{2}\left[ m_{\beta j}\tau _{\beta j}-(m_{\beta
j}+\delta m_{\beta j})\tau _{\beta j}\right] +c^{2}m_{\beta D}\left( \tau
-\tau _{\beta D}\right)  \nonumber \\
&=&\sum_{j=1}^{N\ }c^{2}\delta m_{\beta j}\tau _{\beta j}+c^{2}m_{\beta
D}\left( \tau -\tau _{\beta D}\right)
\end{eqnarray}%
We have therefore in 5D:%
\begin{eqnarray}
\delta \phi (q) &=&\sum_{j=1}^{N}\left( \tilde{k}_{\beta j}q_{\beta j}-%
\tilde{k}_{\alpha j}q_{\alpha j}\right) -\left( \omega _{\beta j}-\omega
_{\alpha j}\right) t_{j}+\left( \varphi _{\beta j}-\varphi _{\alpha j}\right)
\nonumber \\
&&+\left[ \tilde{p}_{\beta ,D}\left( q-q_{\beta ,D}\right) -\tilde{p}%
_{\alpha ,D}\left( q-q_{\alpha ,D}\right) \right] /\hbar
\end{eqnarray}%
where $p_{\beta j},$ $\hbar k_{\beta j}$ and $q_{\beta j}$ have now an
additional 4-component equal respectively to $m_{\beta j}c,$ $\delta
m_{\beta j}c$ and $c\tau _{\beta j}$. For Hermite-Gauss wave packets, this
phase should be evaluated at the mid-point (mid-point theorem \cite%
{metrologia}) $q=\left( q_{\beta ,D}+q_{\alpha ,D}\right) /2$ . This
mid-point phase shift is:%
\begin{eqnarray}
\delta \phi (\left( q_{\beta ,D}+q_{\alpha ,D}\right) /2)
&=&\sum_{j=1}^{N}\left( \tilde{k}_{\beta j}q_{\beta j}-\tilde{k}_{\alpha
j}q_{\alpha j}\right) -\left( \omega _{\beta j}-\omega _{\alpha j}\right)
t_{j}+\left( \varphi _{\beta j}-\varphi _{\alpha j}\right)  \nonumber \\
&&+\left[ (\tilde{p}_{\beta ,D}+\tilde{p}_{\alpha ,D})\left( q_{\alpha
,D}-q_{\beta ,D}\right) /2\right] /\hbar  \label{phase}
\end{eqnarray}%
If energy is conserved, we may use the conservation of the Lagrange
invariant (derived from Stokes theorem):%
\begin{equation}
\left( \tilde{p}_{\alpha j+1}+\tilde{p}_{\beta j+1}\right) \left( q_{\beta
j+1}-q_{\alpha j+1}\right) -\left[ \left( \tilde{p}_{\alpha j}+\tilde{p}%
_{\beta j}\right) +\hbar \left( \tilde{k}_{\beta j}+\tilde{k}_{\alpha
j}\right) \right] \left( q_{\beta j}-q_{\alpha j}\right) =0
\end{equation}%
and obtain the 5D scalar product:

\begin{equation}
\delta \phi (\left( q_{\beta N+1}+q_{\alpha N+1}\right) /2)=\sum_{j=1}^{N} 
\left[ \frac{\tilde{k}_{\beta j}-\tilde{k}_{\alpha j}}{2}\left( q_{\beta
j}+q_{\alpha j}\right) \right] -\left( \omega _{\beta j}-\omega _{\alpha
j}\right) t_{j}+\left( \varphi _{\beta j}-\varphi _{\alpha j}\right)
\end{equation}%
(we have also assumed $q_{\beta 1}=q_{\alpha 1}$)\bigskip . If energy is not
conserved, we may use instead the symplectic Lagrange-Helmholtz invariant in
the quadratic approximation (Hamiltonian of degree 2 at most in position and
momentum):%
\begin{equation}
\frac{\tilde{p}_{\alpha j+1}}{m_{\alpha }^{\ast }}\left( q_{\beta
j+1}-q_{\alpha j+1}\right) -\frac{\tilde{p}_{\alpha j}}{m_{\alpha }^{\ast }}%
\left( q_{\beta j}-q_{\alpha j}\right) =\frac{\tilde{p}_{\beta j+1}}{%
m_{\beta }^{\ast }}\left( q_{\alpha j+1}-q_{\beta j+1}\right) -\frac{\tilde{p%
}_{\beta j}}{m_{\beta }^{\ast }}\left( q_{\alpha j}-q_{\beta j}\right)
\end{equation}%
which reduces to the previous Lagrange invariant with a good approximation
for small relative energy changes. This explains the cancellation of the
action and of the mid-point phase shift in the usual 4D approach as
emphasized in reference \cite{Borde08}.

For the illustration, let us apply the previous formulas to the Bord\'{e}%
-Ramsey interferometer \cite{Borde16,Borde2} represented on the figure.

\FRAME{ftbpF}{7.0369in}{5.2827in}{0pt}{}{}{interfbrcorr.jpg}{\special%
{language "Scientific Word";type "GRAPHIC";maintain-aspect-ratio
TRUE;display "USEDEF";valid_file "F";width 7.0369in;height 5.2827in;depth
0pt;original-width 9.5998in;original-height 7.1996in;cropleft "0";croptop
"1";cropright "1";cropbottom "0";filename
'../BadHonnefBordé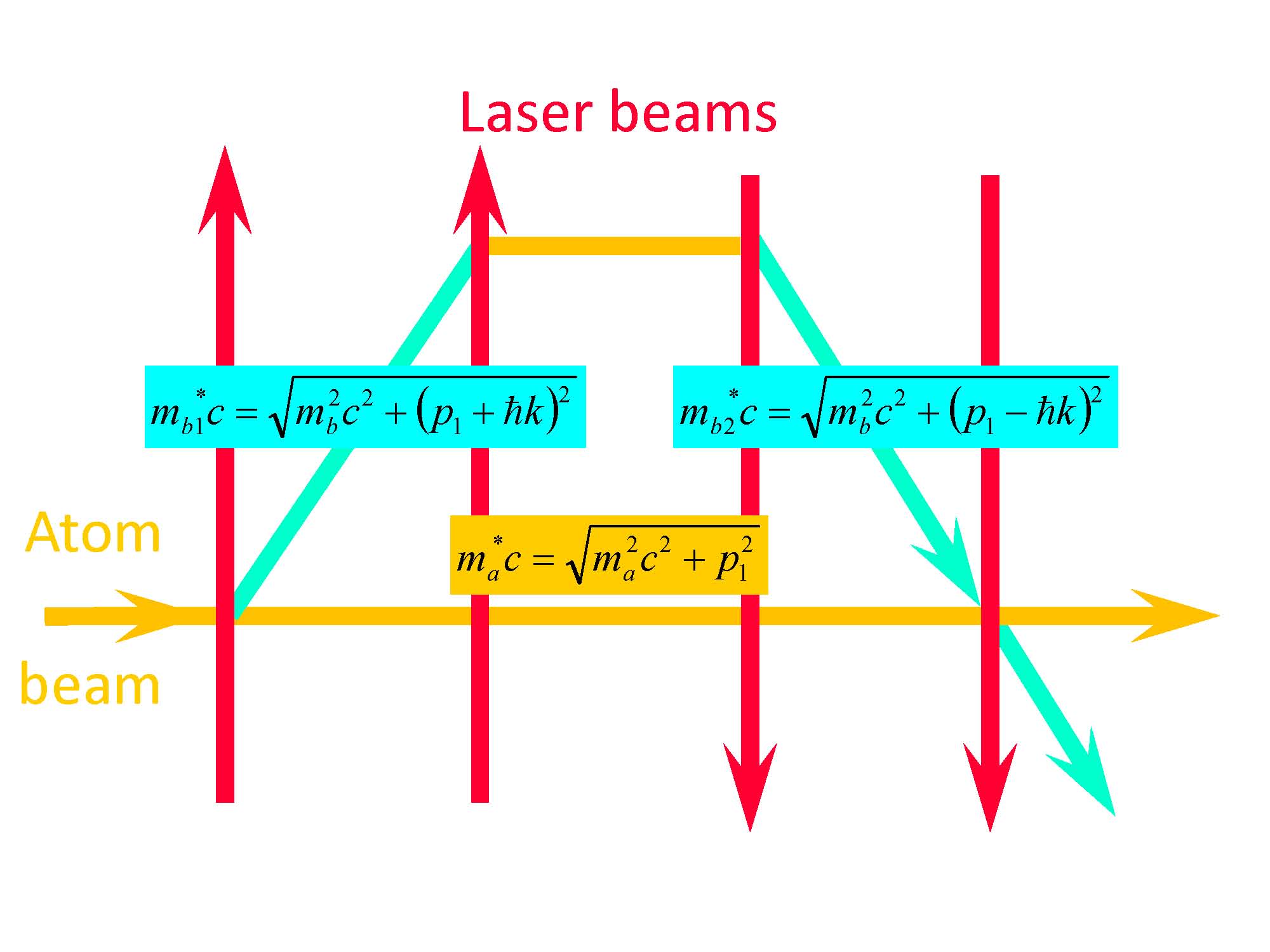';file-properties "XNPEU";}}Formula (\ref%
{phase}) gives for the mid-point 5D phase: 
\begin{eqnarray}
\delta \phi (\left( \widehat{q}_{\beta 4}+\widehat{q}_{\alpha 4}\right) /2)
&=&\vec{k}.\vec{q}_{1}+(m_{b}-m_{a})c^{2}\tau _{1}/\hbar -\omega t_{1} 
\nonumber \\
&&-\vec{k}.\vec{q}_{\beta 2}+(-m_{b}+m_{a})c^{2}\tau _{\beta 2}/\hbar
+\omega t_{2}  \nonumber \\
&&-\vec{k}.\vec{q}_{\beta 3}+(m_{b}-m_{a})c^{2}\tau _{\beta 3}/\hbar -\omega
t_{3}  \nonumber \\
&&+\vec{k}.\vec{q}_{\beta 4}+(-m_{b}+m_{a})c^{2}\tau _{\beta 4}/\hbar
+\omega t_{4}  \nonumber \\
&&+\sum_{j=1}^{4}\left( \varphi _{\beta j}-\varphi _{\alpha j}\right) 
\nonumber \\
&&+\left[ (\text{ }\vec{p}_{\beta b4}+\vec{p}_{\alpha a4}+\hbar \vec{k}%
).\left( \vec{q}_{\alpha 4}-\vec{q}_{\beta 4}\right) /2\right] /\hbar 
\nonumber \\
&&+\left[ (m_{b}+m_{a}+m_{a}-m_{b})\left( \tau _{\alpha 4}-\tau _{\beta
4}\right) /2\right] c^{2}/\hbar
\end{eqnarray}%
In the absence of gravito-inertial fields (e.g. in the inertial frame of the
atoms):%
\begin{eqnarray}
t_{2} &=&t_{1}+T;\text{ \ }\vec{q}_{\beta 2}=\vec{q}_{1}+\frac{\left( \vec{p}%
_{1}+\hbar \vec{k}\right) T}{m_{b1}^{\ast }};\text{ \ }\tau _{\beta 2}=\tau
_{1}+\frac{m_{b}}{m_{b1}^{\ast }}T;\text{ \ }\vec{p}_{\beta 2}=\vec{p}%
_{1}+\hbar \vec{k}  \nonumber \\
t_{3} &=&t_{2}+T^{\prime };\text{ \ }\vec{q}_{\beta 3}=\vec{q}_{1}+\frac{%
\left( \vec{p}_{1}+\hbar \vec{k}\right) T}{m_{b1}^{\ast }}+\frac{\vec{p}%
_{1}T^{\prime }}{m_{a}^{\ast }};\text{ \ }\tau _{\beta 3}=\tau _{1}+\frac{%
m_{b}}{m_{b1}^{\ast }}T+\frac{m_{a}}{m_{a}^{\ast }}T^{\prime };\text{ \ }%
\vec{p}_{\beta b3}=\vec{p}_{1}  \nonumber \\
t_{4} &=&t_{3}+T;\text{ \ }\vec{q}_{\beta 4}=\vec{q}_{1}+\frac{\left( \vec{p}%
_{1}+\hbar \vec{k}\right) T}{m_{b1}^{\ast }}+\frac{\left( \vec{p}_{1}-\hbar 
\vec{k}\right) T}{m_{b2}^{\ast }}+\frac{\vec{p}_{1}T^{\prime }}{m_{a}^{\ast }%
};\text{ }  \nonumber \\
\tau _{\beta 4} &=&\tau _{1}+\frac{m_{b}}{m_{b1}^{\ast }}T+\frac{m_{b}}{%
m_{b2}^{\ast }}T+\frac{m_{a}}{m_{a}^{\ast }}T^{\prime };\text{ \ }\vec{p}%
_{\beta b4}=\vec{p}_{1}-\hbar \vec{k}
\end{eqnarray}%
and for the lower branch:%
\begin{eqnarray*}
\vec{q}_{\alpha 2} &=&\vec{q}_{1}+\frac{\vec{p}_{1}T}{m_{a}^{\ast }};\text{
\ }\tau _{\alpha 2}=\tau _{1}+\frac{m_{a}}{m_{a}^{\ast }}T;\text{ \ }\vec{p}%
_{\alpha 2}=\vec{p}_{1} \\
\vec{q}_{\alpha 3} &=&\vec{q}_{1}+\frac{\vec{p}_{1}\left( T+T^{\prime
}\right) }{m_{a}^{\ast }};\text{ \ }\tau _{\alpha 3}=\tau _{1}+\frac{m_{a}}{%
m_{a}^{\ast }}\left( T+T^{\prime }\right) ;\text{ \ }\vec{p}_{\alpha 3}=\vec{%
p}_{1} \\
\vec{q}_{\alpha 4} &=&\vec{q}_{1}+\frac{\vec{p}_{1}\left( 2T+T^{\prime
}\right) }{m_{a}^{\ast }};\text{ \ }\tau _{\alpha 4}=\tau _{1}+\frac{m_{a}}{%
m_{a}^{\ast }}\left( 2T+T^{\prime }\right) ;\text{ \ }\vec{p}_{\alpha 4}=%
\vec{p}_{1}
\end{eqnarray*}%
We see that the final positions and proper times differ on both arms by
small quantities, owing to the relativistic differences of velocities on
both arms:%
\begin{eqnarray*}
\text{\ }\vec{q}_{\beta 4}-\vec{q}_{\alpha 4} &=&\left( \frac{1}{%
m_{b1}^{\ast }}+\frac{1}{m_{b2}^{\ast }}-\frac{2}{m_{a}^{\ast }}\right) \vec{%
p}_{1}T+\left( \frac{1}{m_{b1}^{\ast }}-\frac{1}{m_{b2}^{\ast }}\right)
\hbar \vec{k}T \\
\tau _{\beta 4}-\tau _{\alpha 4} &=&\left( \frac{m_{b}}{m_{b1}^{\ast }}+%
\frac{m_{b}}{m_{b2}^{\ast }}-\frac{2m_{a}}{m_{a}^{\ast }}\right) T
\end{eqnarray*}%
Finally%
\begin{equation}
\delta \phi =\left[ 2\omega -\left( m_{b1}^{\ast }+m_{b2}^{\ast
}-2m_{a}^{\ast }\right) c^{2}/\hbar \right] T  \label{exact}
\end{equation}%
For each segment, we check the conservation of the symplectic invariant
(Lagrange-Helmoltz), e.g.:%
\begin{equation}
\left( \frac{\text{\ }\vec{p}_{\beta 2}}{m_{b1}^{\ast }}+\frac{\text{\ }\vec{%
p}_{\alpha 2}}{m_{a}^{\ast }}\right) .\left( \vec{q}_{\alpha 2}-\vec{q}%
_{\beta 2}\right) /2+\left( \frac{\text{\ }m_{b}}{m_{b1}^{\ast }}+\frac{%
\text{\ }m_{a}}{m_{a}^{\ast }}\right) \left( \tau _{\alpha 2}-\tau _{\beta
2}\right) c^{2}/2=0
\end{equation}%
\begin{equation}
\frac{\vec{p}_{\alpha 3}+\vec{p}_{\beta 3}}{m_{a}^{\ast }}.\left( \vec{q}%
_{\beta 3}-\vec{q}_{\alpha 3}\right) -\frac{\vec{p}_{\alpha 2}+\text{\ }\vec{%
p}_{\beta 2}}{m_{a}^{\ast }}.\left( \vec{q}_{\beta 2}-\vec{q}_{\alpha
2}\right) +\frac{\text{\ }2m_{a}}{m_{a}^{\ast }}\left( \tau _{\beta 3}-\tau
_{\alpha 3}\right) c^{2}/2-\frac{\text{\ }2m_{a}}{m_{a}^{\ast }}\left( \tau
_{\beta 2}-\tau _{\alpha 2}\right) c^{2}/2=0
\end{equation}%
which correspond to a conserved Lagrange invariant only in the approximation
of conserved energy along the arms of the interferometer. Within this
approximation, we may use the approximate expression%
\begin{eqnarray}
\delta \phi &=&\vec{k}.\left[ \vec{q}_{1}-\left( \vec{q}_{\alpha 2}+\vec{q}%
_{\beta 2}\right) /2-\left( \vec{q}_{\alpha 3}+\vec{q}_{\beta 3}\right)
/2+\left( \vec{q}_{\alpha 4}+\vec{q}_{\beta 4}\right) /2\right]  \nonumber \\
&&+\omega _{ba}\left[ \tau _{1}-\left( \tau _{\alpha 2}+\tau _{\beta
2}\right) /2-\left( \tau _{\alpha 3}+\tau _{\beta 3}\right) /2+\left( \tau
_{\alpha 4}+\tau _{\beta 4}\right) /2\right]  \nonumber \\
&&-\omega \left( t_{1}-t_{2}+t_{3}-t_{4}\right)  \nonumber \\
&=&2\omega T-\frac{\omega _{ba}T}{2}\left( 2\frac{m_{a}}{m_{a}^{\ast }}+%
\frac{m_{b}}{m_{b1}^{\ast }}+\frac{m_{b}}{m_{b2}^{\ast }}\right)  \nonumber
\\
&&-\frac{\hbar k^{2}T}{2}\left( \frac{1}{m_{b1}^{\ast }}+\frac{1}{%
m_{b2}^{\ast }}\right) +\frac{\vec{k}.\vec{p}_{1}T}{2}\left( \frac{1}{%
m_{b2}^{\ast }}-\frac{1}{m_{b1}^{\ast }}\right)
\end{eqnarray}

The same recoil and second-order Doppler corrections are obtained from this
approximate formula to first-order but only expression (\ref{exact}) is
exact. The second-term corresponds to the difference in proper times for the
clock term (that we have called a quantum Langevin twin paradox in reference 
\cite{Borde08}). Note that this clock term implies that the mass differs on
both arms and does not correspond to a different clock on each arm as in the
classical Langevin twin paradox. The coherent quantum superposition of both
arms is essential to generate a clock. The remaining piece of the recoil
shift comes from first-order Doppler shifts contained in the third term and
originating from the spatial $\vec{q}$ part of the shift.

\bigskip

\section{Conclusion}

\textbf{As a conclusion, the motion of massive or massless particles in 5D
follows a null geodesic just as it is the case for photons in 4D. The
Lagrangian is proportional to the interval squared and both vanish for the
real motion. This has the consequence that the phase, which is proportional
to the 5D superaction, will also vanish between two points of the real
trajectory of the particle. As a consequence the phase shift in atom
interferometers results only from the phase jumps introduced by the beam
splitters. }

\section{\protect\bigskip Appendix A}

If we consider an object (such as a clock, a molecule...) composed of a
number of subparticles, the 5D superaction differential is given by the sum:%
\begin{equation}
d\widehat{S}=\sum_{A}\left( -p_{A\mu }dx_{A}^{\mu }+m_{A}c^{2}d\tau
_{A}\right)
\end{equation}%
where $m_{A}$ is the mass of particle $A$. With the following change of
coordinates: 
\begin{equation}
dx_{A}^{\mu }=dX^{\mu }+d\xi _{A}^{\mu }
\end{equation}%
\begin{equation}
d\widehat{S}=-P_{\mu }dX^{\mu }+\sum_{A}\left( -p_{A\mu }d\xi _{A}^{\mu
}+m_{A}c^{2}d\tau _{A}\right)
\end{equation}%
with%
\begin{equation}
P_{\mu }=\sum_{A}p_{A\mu }\text{ and }p_{A\mu }=\left( m_{A}^{\ast }/m^{\ast
}\right) P_{\mu }+\pi _{A\mu }
\end{equation}%
The coordinates $X^{\mu }$ and $\xi _{A}^{\mu }$ are such that%
\begin{equation}
\sum_{A}m_{A}^{\ast }d\xi _{A}^{\mu }=0\text{ and }\xi _{A}^{0}=0
\end{equation}%
(common time coordinate for all the particles of the composed object). One
obtains for the full object: 
\begin{equation}
d\widehat{S}=-P_{\mu }dX^{\mu }+Mc^{2}d\theta =-P_{\hat{\mu}}dX^{\hat{\mu}}=0
\end{equation}%
provided that:%
\begin{eqnarray}
Mc^{2}d\theta &=&\sum_{A}\left( -p_{A\mu }d\xi _{A}^{\mu }+m_{A}c^{2}d\tau
_{A}\right) \\
&=&\sum_{A}\left( -\pi _{Aj}d\xi _{A}^{j}+m_{A}c^{2}d\tau _{A}\right)
\end{eqnarray}%
The source of the proper time $\theta $\ for the object lies in the internal
degrees of freedom and its mass $Mc^{2}$ is given by its internal
Hamiltonian. A well-defined quantum phase for the composed object requires
that it should be in an eigenstate of this internal Hamiltonian.

\section{\protect\bigskip Acknowledgements}

Special thanks to Dr Luc Blanchet and to Dr Peter Wolf for many fruitful and
stimulating discussions.

\end{document}